\documentclass[10pt,twocolumn,twoside,bookmarks=false]{osajnlnt}
\setboolean{shortarticle}{true}
\journal{ol}

\usepackage{endnotes}
\usepackage{pdfpages}
\usepackage{ifthen}
\usepackage{graphicx}
\usepackage{fp}
\usepackage{etex}
\usepackage{subcaption}
\usepackage{booktabs}
\usepackage{tikz}     % Required For theme to work
\usepackage{tikz-3dplot}
\usepackage{import}
\usepackage{wrapfig}
\usepackage[numbered,framed]{matlab-prettifier}

\usepackage{animate}
\usepackage{pgfplots}
\usepackage[english]{babel}
\usepackage{amsmath}
\usepackage{amssymb}
\usepackage{multirow}
\usepackage{float}
\usepackage{epsfig}
\usepackage{rotating}
\usepackage{textcomp}
\usepackage{tyo_tools}

\usepackage[capitalize]{cleveref}

\pgfplotsset{compat=newest}

\definecolor{ASLbot}{RGB}{255,215,0}
\definecolor{ASLtop}{RGB}{255,255,255}
\definecolor{ASLbft}{RGB}{215,219,223}
\definecolor{ASLred}{RGB}{210,16,68}
\definecolor{ASLblu}{RGB}{0,60,124}
\definecolor{ASLgrn}{RGB}{0,141,127}
\definecolor{ASLyel}{RGB}{255,209,139}
\definecolor{ASLyel2}{RGB}{255,153,0}
\definecolor{ASLpurp}{RGB}{101,66,138}
\definecolor{ASLgold}{RGB}{255,205,0}

\definecolor{Line1}{RGB}{68,1,84}
\definecolor{Line2}{RGB}{72,40,110}
\definecolor{Line3}{RGB}{62,74,137}
\definecolor{Line4}{RGB}{49,104,142}
\definecolor{Line5}{RGB}{38,131,142}
\definecolor{Line6}{RGB}{31,157,137}
\definecolor{Line7}{RGB}{53,183,121}
\definecolor{Line8}{RGB}{108,206,89}
\definecolor{Line9}{RGB}{180,221,44}
\definecolor{Line10}{RGB}{253,231,37}

\definecolor{CFAblue}{RGB}{51,0,255}
\definecolor{CFAgreen1}{RGB}{0,255,0}
\definecolor{CFAgreen2}{RGB}{195,255,0}
\definecolor{CFAorange}{RGB}{255,190,0}
\definecolor{CFAred1}{RGB}{255,0,0}
\definecolor{CFAred2}{RGB}{97,0,0}

\definecolor{UNSWyellow}{RGB}{255,215,0}

\definecolor{UNSWred}{RGB}{217,72,38}
\definecolor{UNSWred1}{RGB}{227,33,25}
\definecolor{UNSWred2}{RGB}{193,32,23}
\definecolor{UNSWred3}{RGB}{158,29,21}
\definecolor{UNSWred4}{RGB}{121,25,16}

\definecolor{UNSWblue1}{RGB}{0,174,239}

\definecolor{UNSWpurple1}{RGB}{79,45,127}
\definecolor{UNSWpurple2}{RGB}{107,72,157}
\definecolor{UNSWpurple3}{RGB}{131,99,170}
\definecolor{UNSWpurple4}{RGB}{158,126,185}

\definecolor{UNSWgeneral1}{RGB}{249,178,0}
\definecolor{UNSWgeneral2}{RGB}{255,230,160}
\definecolor{UNSWgeneral3}{RGB}{255,246,254}
\definecolor{UNSWgeneral3}{RGB}{255,244,207}
\usetikzlibrary{fixedpointarithmetic}
\usetikzlibrary{decorations,decorations.pathreplacing,decorations.text,decorations.markings,shapes,shapes.gates.logic.US,arrows,fadings,calc,external,bending}

\makeatletter 
\newcommand{\boxedB}[1]{\textcolor{white}{\fbox{\m@th$\displaystyle#1$}}}
\pgfkeys{/pgf/decoration/.cd, 
	start color/.store in=\startcolor, 
	end color/.store in=\endcolor 
} 

\pgfdeclaredecoration{color change arrow}{initial}{ 
	\state{initial}[width=0pt, next state=line, persistent precomputation={% 
		\pgfmathdivide{50}{\pgfdecoratedpathlength}% 
		\let\increment=\pgfmathresult% 
		\def\x{0}% 
	}]{} 
	\state{line}[width=0.5pt, switch if less than=5.0\pgflinewidth to final, 
	persistent postcomputation={% 
		\pgfmathadd@{\x}{\increment}% 
		\let\x=\pgfmathresult% 
	}]{% 
	\pgfsetlinewidth{\pgflinewidth}% 
	\pgfsetarrows{-}% 
	\pgfpathmoveto{\pgfpointorigin}% 
	\pgfpathlineto{\pgfqpoint{0.75pt}{0pt}}% 
	\pgfsetstrokecolor{\endcolor!\x!\startcolor}% 
	\pgfusepath{stroke}% 
} 
\state{final}{% 
	\pgfsetlinewidth{\pgflinewidth}% 
	\pgfpathmoveto{\pgfpointorigin}% 
	\pgfpathlineto{\pgfpointdecoratedpathlast\advance\pgf@x by2.0\pgflinewidth}% 
	\color{\endcolor!\x!\startcolor}% 
	\pgfusepath{stroke}%	
} 
} 

\makeatother

\numberwithin{equation}{section}

\allowdisplaybreaks

\newlength\fswidth
\newlength\fsheight
\newlength\fCVUwidth
\newlength\fCVUheight
\def\mScale{0.23}
\def\mfps{15}
\newcolumntype{C}[1]{>{\centering\arraybackslash}m{#1}}
\newcolumntype{L}[1]{>{\raggedleft\arraybackslash}m{#1}}
\newcolumntype{R}[1]{>{\raggedright\arraybackslash}m{#1}}
\title{Statistical scene generation for polarimetric imaging systems}

\author[1]{Israel J. Vaughn and Andrey S. Alenin and J. Scott Tyo}

\affil[1]{School of Engineering and IT, University of New South Wales Canberra, Australia}
\affil[*]{Corresponding author: israel.vaughn@gmail.com}

\dates{Compiled \today}
\ociscodes{}
\doi{}
\begin{abstract}
Little publicly available data exists for polarimetric measurements. When designing task specific polarimetric systems, the statistical properties of the task specific data becomes important. Until better polarimetric datasets are available to deduce statistics from, the statistics must be simulated to test instrument performance. Most imaged scenes have been shown to follow a power law power spectral density distribution, for both natural and city scenes. Furthermore, imaged data appears to follow a power law power spectral distribution temporally. We are interested in generating image sets which change over time, and at the same time are correlated between different components (spectral or polarimetric). In this brief communication, we present a framework and provide code to generate such data.
\end{abstract}
\setboolean{displaycopyright}{false}
\begin{document}

\maketitle
\thispagestyle{fancy}
\ifthenelse{\boolean{shortarticle}}{\abscontent}{}
\section{Introduction}
Natural and city images have been shown to have power spectral densities which follow a 
\begin{align}
\frac{A}{|f|^\gamma}\label{sec:intro:eqn:powLaw}
\end{align}
\begin{figure*}
	\centering
	\begin{align}
		x_{n_1,n_2,n_3} =
		\frac{1}{N_1N_2N_3}\sum_{k_1=0}^{N_1-1}\left(\sum_{k_2=0}^{N_2-1}\left(\sum_{k_3=0}^{N_3-1}e^{2\pi i (n_1k_1/N_1 + n_2k_2/N_2 + n_3k_3/N_3)}F(k_1,k_2,k_3)X_{k_1,k_2,k_3}\right)\right)\label{sec:theory:eqn:iDFT}
	\end{align}
%	\begin{align}
%	x_{n_1,n_2,n_3}& =
%	\frac{1}{N_1N_2N_3}\sum_{k_1=0}^{N_1-1}\left(\sum_{k_2=0}^{N_2-1}\left(\sum_{k_3=0}^{N_3-1}e^{2\pi i (n_1k_1/N_1 + n_2k_2/N_2 + n_3k_3/N_3)}F(k_1,k_2,k_3)X_{k_1,k_2,k_3}\right)\right)\\
%	&=\frac{1}{N_1N_2N_3}\sum_{k_1=0}^{N_1-1}\left(\sum_{k_2=0}^{N_2-1}\left(\sum_{k_3=0}^{N_3-1}\cos\left(2\pi (n_1k_1/N_1 + n_2k_2/N_2 + n_3k_3/N_3)\right)F(k_1,k_2,k_3)X_{k_1,k_2,k_3}\right)\right)\notag\\
%	&+\frac{i}{N_1N_2N_3}\sum_{k_1=0}^{N_1-1}\left(\sum_{k_2=0}^{N_2-1}\left(\sum_{k_3=0}^{N_3-1}\sin\left(2\pi (n_1k_1/N_1 + n_2k_2/N_2 + n_3k_3/N_3)\right)F(k_1,k_2,k_3)X_{k_1,k_2,k_3}\right)\right).
%	\end{align}
\end{figure*}
distribution \cite{dong1995statistics,van1996modelling,torralba2003statistics,srivastava2003advances} for visible color bands. No such empirical study has been undertaken for Mueller or Stokes polarimetric images as there is little to no publicly available data. We are currently working towards collecting and publishing such data, but we require a substitute for instrument design until the empirical study is completed. Our group designs novel polarimetric and spectropolarimetric measuring instruments, with task specific estimation or detection tasks in mind. We recently introduced spectral and polarimetric design frameworks based on bandwidth tradeoffs in the channel space \cite{Vaughn15n1,Vaughn15n2,vaughn2016bounds,alenin2017optimal,vaughn2017focal}, however the analysis is lacking some statistical robustness due to the lack of good measured truth data for test and comparison purposes. In this brief communication we outline how we generate random samples from specific power spectral density functions, including correlated components, and present the algorithm to produce these correlated samples. Matlab code is available in the ancillary materials under a GPLv3 license. 
\section{Theory}
Statistical distributions \cite{childers1978modern,deodatis1996simulation,broersen2003generating,ashby2011discrete} can be generated via a variety of methods. We use the convolution/Fourier transform method here \cite{kasdin1995discrete,ashby2011discrete}, whereby some white noise distribution is generated, then filtered via convolution or multiplication in the Fourier domain. This method is straightforward to implement and results in fast computational times. The noise can also be correlated, correlation of Gaussian white noise is straightforward to compute and generate \cite{Barrett04}, however this does not necessarily translate into the easily generated correlated noise with a specific covariance matrix and a specific distribution \cite{broersen2003generating}. In this communication we present algorithms to generate samples of specific PSD distributions with a specific covariance matrix. Although the covariance matrix between polarimetric images is currently unknown, we can set reasonable values for polarimetric imaging system evaluation. Polarimetric scenes will have specific correlations between $s_0, s_1, s_2, s_3$, for different environments, however these correlations are still being determined. We intend to use large polarimetric datasets in the future to determine these correlations, and at that time we will refactor the sample generation code to represent a proper covariance matrix. 

In this communication we present 3 algorithms; 1) generation of a number of images which obey a power law PSD spatially, but are temporally uncorrelated, 2) generation of a number of images which obey a power law PSD spatio-temporally, 3) generation of $N$ sets of images which obey a power law PSD spatio-temporally and are correlated between the sets. The power law PSDs are generated in the following way:
\begin{itemize}
	\item images or sets of images of white noise are generated.
	\item the images are are taken to the frequency domain using a 2-dimensional or 3-dimensional fast Fourier transform (FFT) depending on the desire for a temporal power law PSD.
	\item Fourier domain filters are generated by using the power law in \cref{sec:intro:eqn:powLaw} where $f = \sqrt{\xi^2 + \eta^2}$ or $f = \sqrt{\xi^2 + \eta^2 + \nu^2}$ and $\xi$ corresponds to the spatial frequencies in the $x$-direction, $\eta$ corresponds to the spatial frequencies in the $y$-direction, and $\nu$ corresponds to the temporal frequencies.
\end{itemize}
\subsection{Mean and variance}
The Fourier transform of normally distributed Gaussian noise is again normally distributed Gaussian noise \cite{oberhettinger2014fourier}. Once filtered by a power law PSD, the spatio-temporal data will have a specific mean and variance, which we would like to derive analytically and allow a user to specify. We present the derivation for white Gaussian noise (WGN) under a discrete Fourier transform (DFT) here. In the algorithms presented in this communication, we generate zero-mean WGN in the spatio-temporal domain, then transform to the Fourier domain for filtering. This results in zero-mean WGN in the discrete Fourier domain, with an associated Hermiticity condition. Each spatio-temporal frequency location can be treated as a random variable, $X_{k_1,k_2,k_3}$ at location $(k_1,k_2,k_3)$ of the DFT domain. The inverse discrete Fourier transform is defined in \cref{sec:theory:eqn:iDFT} at the point $x_{n_1,n_2,n_3}$.
%where we have used the vector summation notation, $\vect{k} = (\xi,\eta,\nu)$, $\vect{N-1} = (N_1-1, N_2-1,N_3-1)$, $\vect{k}/\vect{N}$ is defined as the elementwise quantity $(\xi/N_1,\eta/N_2,\nu/N_3)$
We can then compute the mean and variance of $x_{n_1,n_2,n_3}$ assuming that the random variables $X_{k_1,k_2,k_3}$ are independent with variance $\sigma^2$ and $F(k_1,k_2,k_3)$ is the Fourier domain filtering function. If we assume that the mean of each $X_{k_1,k_2,k_3}$ is $0$, then the mean of $x_{n_1,n_2,n_3}$ is also $0$. The variance for $x_{n_1,n_2,n_3}$ is then 
\begin{align}
	&\var(x_{n_1,n_2,n_3}) =\notag\\ 
	&\frac{1}{(N_1N_2N_3)^2}\sum_{k_1=0}^{N_1-1}\left(\sum_{k_2=0}^{N_2-1}\left(\sum_{k_3=0}^{N_3-1}F^2(k_1,k_2,k_3)\sigma^2\right)\right)\label{sec:theory:eqn:var}.
\end{align}
Notice the lack of dependence on $(k_1,k_2,k_3)$ due to the white noise assumption. Additionally if zero mean WGN with a variance of $1$ is used as the initial spatio-temporal input, then the DFT will have a variance of $\sigma^2 = N_1N_2N_3$ for the standard FFT definition. This allows us to analytically adjust the variance and mean of the generated datasets to specified values in our generation algorithms. Note that for a spatial distribution only \cref{sec:theory:eqn:var} reduces to the 2-dimensional case.\
\subsection{Correlation}
Stokes images can be thought of as a set of 4 spatio-temporal data cubes which are correlated. The third algorithm in the following section can generate $K$ correlated spatio-temporal datacubes as samples from a specific 3-dimensional power law PSD. $\Sigma$ is the input $K\times K$ covariance matrix. 

There is some subtlety involved in both specifying $\Sigma$ and the mean and STD of each data cube. Given a vector $\vect{X}$ of independent variables with $0$ mean and variance $1$, we have the covariance matrix given by
\begin{align*}
\Sigma_\vect{X} = \mathbb{E}\left[\left(\vect{X}-\mathbb{E}\left[\vect{X}\right]\right)\left(\vect{X}-\mathbb{E}\left[\vect{X}\right]\right)^T\right] = \mathbb{E}\left[\vect{X}\vect{X}^T\right] = I
\end{align*}
where $I$ is the identity matrix. Then if $\vect{Z} = L\vect{X}$ we have
\begin{align}
\Sigma_\vect{Z} &= \mathbb{E}\left[\vect{Z}\vect{Z}^T\right]\notag\\ 
&= \mathbb{E}\left[L\vect{X}(L\vect{X})^T\right]\notag\\
&= \mathbb{E}\left[L\vect{X}\vect{X}^TL^T\right]\notag\\ 
&= L\mathbb{E}\left[\vect{X}\vect{X}^T\right]L^T,\text{by linearity of expectation}\notag\\ 
&= LL^T.\label{sec:theory:eqn:sigmaEqVar}
\end{align}
$L$ can be found from $\Sigma$ via the Cholesky decomposition. In our application we specify $\Sigma_\vect{Z}$, compute $L$, obtain a linear transformation of the datasets using $\vect{Z} = L\vect{X}$, then add the specified means to obtain specific Stokes datasets.
\section{Algorithms and Code}
\begin{algorithm*}
	\caption{genObsPowerLaw. Note that the calculation for $nVar$ does not square the denominator as shown in \cref{sec:theory:eqn:var}, this is due to the factor of $numX\cdot numY$ which is imposed on the variance from the 2-dimensional FFT. \label{alg:genObsPowerLaw}}
	\begin{algorithmic}[1]
		\Function{genObsPowerLaw}{$N,A,\gamma,rA,rB$,\textproc{NFcn},$FP$}
		\State $rI \gets $ \Call{NFcn}{$N$}\Comment{$rI$ is a set of $N$ random images with STD $1$ and mean $0$ when \textproc{NFcn} is WGN}
		\State $F \gets \frac{\sqrt{A}}{f^{\gamma/2}}$\Comment{Take the square root to get the filter for a PSD sample}
		\For{$k=1$ to $N$}
		\State $rI_{FFT}[k] \gets \text{FFT2}\left(rI[k]\right)$\Comment{take 2-D FFT of the current image}
		\State $rIF_{FFT}[k] \gets F\cdot rI_{FFT}[k]$\Comment{Filter the image in the spatial frequency domain}
		\State $rIF[k] \gets \text{IFFT2}\left(rIF_{FFT}[k]\right)$\Comment{take 2-D IFFT of the filtered image}		
		\EndFor\label{genObsEndForOne}
		\State $nVar \gets $\Call{Sum}{$F\odot F$}$/(numX\cdot numY)$\Comment{$\odot$ here denotes elementwise multiplication. \Call{Sum}{} sums over all elements.}
		\State $rIF \gets (rA/\sqrt{nVar})\cdot rIF + rB$\Comment{Specify the STD and mean of $rIF$, first normalizing back to an STD of $1$ via $nVar$.}
		\If{$FP=$ true}
		$rIF[rIF < 0] \gets 0$\Comment{Any value less that zero is truncated to zero}
		\EndIf
		\State \textbf{return} $rIF$\Comment{Random image samples in the spatial domain with the specified PSD}
		\EndFunction
	\end{algorithmic}
\end{algorithm*}
\begin{algorithm*}
	\caption{genObsPowerLaw2. Note that the calculation for $nVar$ does not square the denominator as shown in \cref{sec:theory:eqn:var}, this is due to the factor of $numX\cdot numY\cdot N$ which is imposed on the variance from the 3-dimensional FFT.\label{alg:genObsPowerLaw2}}
	\begin{algorithmic}[1]
		\Function{genObsPowerLaw2}{$N,A,\gamma,rA,rB$,\textproc{NFcn},$FP$}
		\State $rI \gets $ \Call{NFcn}{$N$} \Comment{$rI$ is a set of $N$ random images with STD $1$ and mean $0$ when \textproc{NFcn} is WGN}
		\State $F \gets \frac{\sqrt{A}}{f^{\gamma/2}}$\Comment{Take the square root to get the filter for a PSD sample, $f$ here is 3-D frequency.}
		\State $rI_{FFT} \gets \text{FFT3}\left(rI\right)$\Comment{take 3-D FFT of the current image}
		\State $rIF_{FFT} \gets F\cdot rI_{FFT}$\Comment{Filter the image in the spatio-temporal frequency domain}
		\State $rIF \gets \text{IFFT3}\left(rIF_{FFT}\right)$\Comment{take 3-D IFFT of the filtered image}
		\State $nVar \gets $\Call{Sum}{$F\odot F$}$/(numX\cdot numY\cdot N)$\Comment{$\odot$ here denotes elementwise multiplication. \Call{Sum}{} sums over all elements.}
		\State $rIF \gets (rA/\sqrt{nVar})\cdot rIF + rB$\Comment{Specify the STD and mean of $rIF$, first normalizing back to an STD of $1$ via $nVar$.}
		\If{$FP=$ true}
		$rIF[rIF < 0] \gets 0$\Comment{Any value less that zero is truncated to zero}
		\EndIf
		\State \textbf{return} $rIF$\Comment{Random image samples in the spatio-temporal domain with the specified PSD}
		\EndFunction
	\end{algorithmic}
\end{algorithm*}
\begin{algorithm*}
	\caption{genObsPowerLaw3. Note that the matrix multiplication in line 12 treats the data as a large set of $K\times 1$ vectors and multiplies each vector by $L$. $rA$ can be eliminated because it is specified in the diagonal of $\Sigma$.}\label{alg:genObsPowerLaw3}
	\begin{algorithmic}[1]
		\Function{genObsPowerLaw3}{$N,A,\gamma,rB$,\textproc{NFcn},$FP$,$K$,$\Sigma$}
		\State $L \gets $\Call{Chol}{$\Sigma$}\Comment{Compute the Cholesky decomposition of the covariance matrix and take the upper triangular matrix}
		\For{$k=1$ to $K$}
		\State $rI[k] \gets $ \Call{NFcn}{$N$} \Comment{$rI$ is a set of $K$ random image sets with STD $1$ and mean $0$ when \textproc{NFcn} is WGN}
		\EndFor		
		\State $F \gets \frac{\sqrt{A}}{f^{\gamma/2}}$\Comment{Take the square root to get the filter for a PSD sample, $f$ here is 3-D frequency.}
		\State $nVar \gets $\Call{Sum}{$F\odot F$}$/(numX\cdot numY\cdot N)$\Comment{$\odot$ here denotes elementwise multiplication. \Call{Sum}{} sums over all elements.}
		\For{$k=1$ to $K$}
		\State $rI_{FFT} \gets \text{FFT3}\left(rI[k]\right)$\Comment{take 3-D FFT of the current image}
		\State $rIF_{FFT} \gets F\cdot rI_{FFT}$\Comment{Filter the image in the spatio-temporal frequency domain}
		\State $rIFt \gets \text{IFFT3}\left(rIF_{FFT}\right)$\Comment{take 3-D IFFT of the filtered image}		
		\State $rIF[k] \gets (1/\sqrt{nVar})\cdot rIFt$\Comment{set the variance to 1.}	
		\EndFor
		\State $rI \gets L\cdot rI$\Comment{Matrix-vector multiply $L$ by $rI$ as pages of $K\times 1$ vectors. The datacubes will be correlated as $\Sigma$.}			
		\State $rI \gets rI + rB$\Comment{Add the means. Added as a $K\times 1$ vector to the $K$ pages.}
		\If{$FP=$ true}
		%		$rIF[rIF < 0] \gets 0$\Comment{Any value less that zero is truncated to zero}
		\EndIf
		\State \textbf{return} $rIF$\Comment{Random image samples in the spatio-temporal domain with the specified PSD}
		\EndFunction
	\end{algorithmic}
\end{algorithm*}
We describe the specific algorithms and the associated Matlab code here. The first algorithm generates sets of random images with a spatial PSD given by 
\begin{align}
\frac{A}{|f|^\gamma}
\end{align}
where $f = \sqrt{\xi^2 + \eta^2}$. Note that in the algorithms below, we cannot evaluate the PSD filter for $f = 0$ since that results in an infinite value, so the $f=0$ value is replaced with $f=\varepsilon$ for some $\varepsilon > 0$. We can specify the number of images, $N$, the desired mean parameter $rB$, the desired standard deviation, $rA$, the white noise generating function itself, \textproc{NFcn}, and the force positive parameter $FP$. The $FP$ parameter is for optical image sensors, which have non-negative measurements, if set to true, any negative values are truncated to 0. Note that if \textproc{NFcn} does not generate independent samples then the computed mean and variance may be incorrect.
%\onecolumn
%\lstinputlisting[caption = {PSD generation for temporally independent images.}]{./code/genPowerLawObs.m}
%\twocolumn

The second algorithm generates a set of $N$ images, but assumes that they obey a 3-dimensional power law PSD distribution \cite{dong1995statistics} where $f = \sqrt{\xi^2 + \eta^2 + \nu^2}$. This results in Brownian noise or derivatives of Brownian noise depending on the value of $\gamma$. Statistics of natural images obey a spatial power law PSD where $\gamma \approx 2$ \cite{torralba2003statistics}, with slightly different numbers for different types of images. This implies that we cannot use $\gamma = N$ for our 3-dimensional filtering algorithm, where $N$ is derived empirically for static images. The details about how \begin{figure}[h]
	\centering
	\begin{animateinline}[]{\mfps}
		\multiframe{64}{iStep=1+1}
		{
			\begin{tabular}{cc}				
				$s_{0}$ & $s_{1}$\\[1mm]
				\hline
				\rule{0pt}{0.15\textwidth}\raisebox{-.425\height}{\hspace{-10bp}\includegraphics[width=\mScale\textwidth]{./movies/randObsCubeSample_0\iStep-1.jpg}} & 
				\raisebox{-.425\height}{\includegraphics[width=\mScale\textwidth]{./movies/randObsCubeSample_0\iStep-2.jpg}\hspace{-12bp}}\\[15mm]
				$s_{2}$ & $s_{3}$\\[1mm]
				\hline
				\rule{0pt}{0.15\textwidth}\raisebox{-.425\height}{\hspace{-10bp}\includegraphics[width=\mScale\textwidth]{./movies/randObsCubeSample_0\iStep-3.jpg}} & 
				\raisebox{-.425\height}{\includegraphics[width=\mScale\textwidth]{./movies/randObsCubeSample_0\iStep-4.jpg}\hspace{-12bp}}				
			\end{tabular}
		}
	\end{animateinline}
	\caption{A sample data cube for correlated Stokes parameters using \textproc{genObsPowerLaw3}. The range of digital values for $s_0$ is $[90,310]$ and the range for $s_1,s_2,s_3$ is $[-90,90]$. The images shown here are scaled to $[0,255]$ for $s_0$ and $s_1,s_2,s_3$ respectively. Animated in the electronic version.\label{sec:stokes:fig:randObsCubeEx}}
	\vspace{-20bp}
\end{figure}the exponent in the PSD changes for Brownian motion processes in different Euclidean dimensions is given by Bassingthwaighte and Raymond \cite{bassingthwaighte1995evaluation} and by Heneghan, Lowen, and Teich for the 2-dimensional case \cite{heneghan1996two}. This communication will not delve into details, but the exponent is dependent on $H = E + 1 − D$ where $E$ is the Euclidean dimension, $D$ is the fractal dimension of the process, and $H$ is the Hurst coefficient. This must be taken into account when using the 3-dimensional algorithm to maintain specific spatial PSD distributions. The PSD is proportional to \cite{bassingthwaighte1995evaluation}
\begin{align}
	\frac{A}{|f|^{2H+1}}.
\end{align}
%for the \emph{white noise datacubes}, note that this does not imply that the filtered samples which follow the power law PSD will have the same covariance matrix.
\section{Stokes Data}
In the previous section we presented 3 algorithms which provide samples of power law PSD distributions for a variety of situations. In this section we will use the \textproc{genObsPowerLaw3} algorithm to generate samples as inputs for our analysis of Stokes parameter measuring instruments. We have selected specific $\Sigma,rB$ parameters for\begin{figure}[h]
	\centering
	\begin{animateinline}[]{\mfps}
		\multiframe{64}{iStep=1+1}
		{
			\begin{tabular}{cc}				
				$s_{0}$ & $s_{1}$\\[1mm]
				\hline
				\rule{0pt}{0.15\textwidth}\raisebox{-.425\height}{\hspace{-10bp}\includegraphics[width=\mScale\textwidth]{./movies/randObsCubeSampleUnCorr_0\iStep-1.jpg}} & 
				\raisebox{-.425\height}{\includegraphics[width=\mScale\textwidth]{./movies/randObsCubeSampleUnCorr_0\iStep-2.jpg}\hspace{-12bp}}\\[15mm]
				$s_{2}$ & $s_{3}$\\[1mm]
				\hline
				\rule{0pt}{0.15\textwidth}\raisebox{-.425\height}{\hspace{-10bp}\includegraphics[width=\mScale\textwidth]{./movies/randObsCubeSampleUnCorr_0\iStep-3.jpg}} & 
				\raisebox{-.425\height}{\includegraphics[width=\mScale\textwidth]{./movies/randObsCubeSampleUnCorr_0\iStep-4.jpg}\hspace{-12bp}}				
			\end{tabular}
		}
	\end{animateinline}
	\caption{A sample data cube for uncorrelated Stokes parameters using \textproc{genObsPowerLaw}. The range of digital values for $s_0$ is $[90,310]$ and the range for $s_1,s_2,s_3$ is $[-90,90]$. The images shown here are scaled to $[0,255]$ for $s_0$ and $s_1,s_2,s_3$ respectively. Animated in the electronic version.\label{sec:stokes:fig:randObsCubeUnCorrEx}}
	\vspace{-5bp}
\end{figure} each Stokes parameter.
The covariance matrix is
%\begin{figure}[H]
%	\centering
%	\begin{animateinline}[]{\mfps}
%		\multiframe{1}{iStep=1+1}
%		{
%			\begin{tabular}{cc}				
%				$s_{0}$ & $s_{1}$\\[1mm]
%				\hline
%				\rule{0pt}{0.15\textwidth}\raisebox{-.425\height}{\hspace{-10bp}\includegraphics[width=\mScale\textwidth]{./movies/randObsCubeSampleUnCorr_0\iStep-1.jpg}} & 
%				\raisebox{-.425\height}{\includegraphics[width=\mScale\textwidth]{./movies/randObsCubeSampleUnCorr_0\iStep-2.jpg}\hspace{-12bp}}\\[15mm]
%				$s_{2}$ & $s_{3}$\\[1mm]
%				\hline
%				\rule{0pt}{0.15\textwidth}\raisebox{-.425\height}{\hspace{-10bp}\includegraphics[width=\mScale\textwidth]{./movies/randObsCubeSampleUnCorr_0\iStep-3.jpg}} & 
%				\raisebox{-.425\height}{\includegraphics[width=\mScale\textwidth]{./movies/randObsCubeSampleUnCorr_0\iStep-4.jpg}\hspace{-12bp}}				
%			\end{tabular}
%		}
%	\end{animateinline}
%	\caption{A sample data cube for uncorrelated Stokes parameters using \textproc{genObsPowerLaw}. The range of digital values for $s_0$ is $[90,310]$ and the range for $s_1,s_2,s_3$ is $[-90,90]$. The images shown here are scaled to $[0,255]$ for $s_0$ and $s_1,s_2,s_3$ respectively. Animated in the electronic version.\label{sec:stokes:fig:randObsCubeUnCorrEx}}
%\end{figure}
\begin{align}
\Sigma = \begin{bmatrix}
500 &  200 &  200 & 100\\
200 &  350 & -100 & 50\\
200 & -100 &  350 & 50\\
100 &   50 &   50 & 200
\end{bmatrix}
\end{align}
and
\begin{align}
rB = \begin{bmatrix}
200\\0\\0\\0
\end{bmatrix}
\text{ corresponding to }
\begin{bmatrix}
s_0\\s_1\\s_2\\s_3
\end{bmatrix}.
\end{align}
We set $\gamma = 3$,$A=1$, and the frequency ranges to $\xi,\eta \in [-1,1];\;\nu\in[-1/8,1/8]$ and the datacube size to $512\times512\times64$, i.e., $64$ temporal images of size $512\times512$. We set $\gamma=3$ instead of $\gamma=2$ here because we are generating spatio-temporal images and using a 3-dimensional filter \cite{bassingthwaighte1995evaluation}, $\gamma = 2$ may be used to generate sets of images with only 2-dimensional filtering via  \textproc{genObsPowerLaw}, with any sets of images being temporally uncorrelated. See \cref{sec:stokes:fig:randObsCubeEx} for the spatio-temporal Stokes correlated images and \cref{sec:stokes:fig:randObsCubeUnCorrEx} for an example of images from \textproc{genObsPowerLaw} with $\gamma=2$.

\begin{figure}[h]
	\centering
	% This file was created by matlab2tikz.
%
%The latest updates can be retrieved from
%  http://www.mathworks.com/matlabcentral/fileexchange/22022-matlab2tikz-matlab2tikz
%where you can also make suggestions and rate matlab2tikz.
%
%
\begin{tikzpicture}

\begin{axis}[%
width=\fCVUwidth,
height=\fCVUheight,
at={(0,0)},
scale only axis,
xmin=0,
xmax=64,
xlabel={sample time},
xmajorgrids,
ymin=-40,
ymax=40,
ylabel={mean difference},
ymajorgrids,
axis background/.style={fill=white},
legend style={legend cell align=left,align=left,draw=white!15!black}
]
\addplot [color=Line8,solid,thick]
  table[row sep=crcr]{%
1	-0.93970248701902\\
2	-1.20527786756734\\
3	-0.794401344558216\\
4	-0.515709186838498\\
5	0.219030348844287\\
6	0.546682978727256\\
7	0.38906676021005\\
8	0.0277469393170977\\
9	-0.445714497643591\\
10	-0.186223287871154\\
11	1.34109722860468\\
12	0.507780299113667\\
13	-0.0405873286031452\\
14	-0.236313460622796\\
15	-0.139037997702451\\
16	-0.144408529875303\\
17	-0.348791021988163\\
18	-0.372468390353869\\
19	0.405898757952808\\
20	-0.0203238276627626\\
21	0.671020662431998\\
22	1.35386406707777\\
23	0.925957159425149\\
24	1.02208177707146\\
25	1.19448384398093\\
26	1.21870735545763\\
27	1.29640668218415\\
28	0.503038352496304\\
29	0.280500037025008\\
30	-0.517484882962755\\
31	-0.899215108357989\\
32	-1.72078717061675\\
33	-0.354490454083335\\
34	0.930161852072838\\
35	0.977991837524432\\
36	1.32584100914631\\
37	0.433746322817686\\
38	-0.761890608439225\\
39	-1.32397903994278\\
40	-0.952276849296385\\
41	0.262350341385654\\
42	0.364357833435426\\
43	0.241836803893462\\
44	-0.363065421628546\\
45	-0.565763379589343\\
46	-0.498820564378857\\
47	-0.591649694633162\\
48	-0.827209156427307\\
49	-1.10934484694091\\
50	-0.665575800545272\\
51	-0.657794898435997\\
52	-0.454201625428545\\
53	-0.268195517245677\\
54	-0.057567582931499\\
55	-0.0185380751286929\\
56	0.939004145102143\\
57	0.819634730840382\\
58	-0.179159752561619\\
59	-0.321969134725855\\
60	0.24984401820871\\
61	0.415419693443433\\
62	0.77103315102545\\
63	-0.0730249651740246\\
};
\addlegendentry{\textproc{genObsPowerLaw3}};

\addplot [color=Line4,dashed,thick]
  table[row sep=crcr]{%
1	-1.15271621175146\\
2	-0.404159432018711\\
3	4.89288083119214\\
4	-9.90529646751752\\
5	12.1101933369551\\
6	-10.9529089931033\\
7	19.6120990900891\\
8	-32.6861490741245\\
9	10.2345823605\\
10	11.7601112487287\\
11	-5.25145003189048\\
12	1.76984676789944\\
13	-0.872171811392379\\
14	3.52842824281315\\
15	-5.20622054557403\\
16	-5.57598864451263\\
17	-5.08781952907722\\
18	25.1744195746382\\
19	-7.04441364765174\\
20	1.64045251476208\\
21	-11.4539881727889\\
22	22.9537015197699\\
23	-22.8418395778529\\
24	4.32843296629511\\
25	1.67937885309156\\
26	-12.8769144640532\\
27	10.5476736286181\\
28	-3.12624033495995\\
29	10.058752537372\\
30	-4.81066076602583\\
31	4.10713165490398\\
32	-7.46988770668075\\
33	-9.08926176621257\\
34	9.98644181000316\\
35	-4.12966386178102\\
36	9.01687045134124\\
37	-8.34884975346615\\
38	6.27162806136018\\
39	-7.73246035662355\\
40	-0.450286073530129\\
41	-1.4724155906923\\
42	25.5934057110454\\
43	-11.1522084165571\\
44	1.43225459416135\\
45	-9.57984593293249\\
46	-5.24532994003545\\
47	12.3741992888352\\
48	-4.05811300161633\\
49	11.5772795344063\\
50	-17.3897506434488\\
51	7.8258222817876\\
52	-1.15663679773689\\
53	3.76432428106284\\
54	-12.2262386063727\\
55	3.51122062128745\\
56	-11.9909539850707\\
57	20.1305316473484\\
58	-6.1697450871131\\
59	2.23865812074397\\
60	4.97136986081974\\
61	-9.56104547482017\\
62	-0.682336245489505\\
63	12.4903558065239\\
};
\addlegendentry{\textproc{genObsPowerLaw}};

\end{axis}
\end{tikzpicture}%
	\caption{The mean temporal difference conveys the temporal variation differences between \textproc{genObsPowerLaw} and \textproc{genObsPowerLaw2},\textproc{genObsPowerLaw3}. The mean difference is computed at each time step, the figure shows $s_0$ data.\label{sec:stokes:fig:UvC}}
\end{figure}
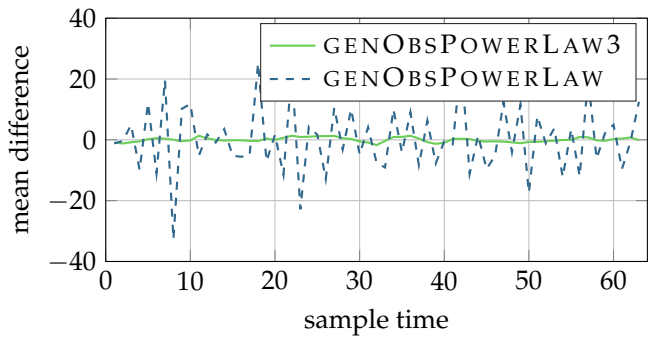

The difference between the spatio-temporally filtered data generated by \textproc{genObsPowerLaw2},\textproc{genObsPowerLaw3} and the spatially filtered data generated by \textproc{genObsPowerLaw} is shown in \cref{sec:stokes:fig:UvC} for the $s_0$ component. The specified means and variances of both data are identical for $512\times512\times64$ image sets as shown in \cref{sec:stokes:fig:randObsCubeEx,sec:stokes:fig:randObsCubeUnCorrEx}. The mean difference is computed as follows
\begin{align}
md(t) = \frac{1}{512^2}\sum_{k=1}^{512} \sum_{n=1}^{512} \bigg[rIF(k,n,t+1)-rIF(k,n,t)\bigg]
\end{align}
where $t=1,2,\cdots,64$ is the discrete temporal variable, and $k,n$ index the spatial locations of the image sets. The mean difference between the two types of data generation is shown in \cref{sec:stokes:fig:UvC}, notice the higher variance of the \textproc{genObsPowerLaw} algorithm. This approach for polarimetric images is also applicable to multi- or hyper-spectral image sets.
\section{Conclusion}
We have derived and presented 3 algorithms for statistical scene generation for polarimetric (or multi- hyper-spectral) image sets. The third algorithm, \textproc{genObsPowerLaw3}, allows for correlation between multiple sets to occur, i.e., correlation between $s_1$ and $s_2$ of a Stokes image or correlation between the red and blue channels of a color image set. The algorithms generate statistically accurate sample images to test imaging system performance, for families of power law power spectral distributions. The specific parameters of the power laws can be specified for polarimetric data once future studies determine the parameters empirically, but a reasonable assumption is to use the parameters similar to those derived from color images. We present an example using our image generation algorithms, and elucidate the difference between spatial only power law distributions and spatio-temporal power law distributions.
%\section{Supplementary Material}
%
%Consult the \href{https://www.osapublishing.org/submit/style/multimedia.cfm}{Author Guidelines for Supplementary Materials in OSA Journals} for details on accepted types of materials and instructions on how to cite them. All materials must be associated with a figure, table, or equation or be referenced in the results section of the manuscript.
%
%2D and 3D image files and video must be labeled “Visualization,” not “Movie,” “Video,” “Figure,” etc. Machine-readable data (for example, csv files) must be labeled  “Data File.”  Number data files and visualizations consecutively, e.g., “Visualization 1, Visualization 2, etc.”. Such items should be mentioned in the text as either “Dataset” or “Code,” as appropriate, and also be cited in the references list. For example, see Dataset 1 (Ref. [1])  and Code 1 (Ref. [2]). Here are two examples:
%
%Sample Dataset Citation
%
%1. T. Ireno and R. Tadaa, "Chemical and mineral compositions of sediments from ODP Site 127-797" (Geological Institute, University of Tokyo, 2009), http://dx.doi.org/10.1594/PANGAEA.726855.
%
%Sample Code Citation
%
%2. Zima Engineering, ZIMA-CAD-Parts: Application for management of CAD files and projects (version 0.5.0-beta1) [software] (2013), http://sourceforge.net/projects/zima-cad-parts/.

\section*{Funding}
Asian Office of Aerospace Research and Development (FA2386-15-1-4098).

%\section*{Acknowledgments}
%
%Formal funding declarations should not be included in the acknowledgments but in a Funding Information section as shown above. The acknowledgments may contain information that is not related to funding:
%
%The authors thank Matthew A. Kupinski for asking questions about MPAs which led to this work.

%\section*{Supplemental Documents}
%\emph{Optica} authors may include supplemental documents with the primary manuscript. For details, see \href{http://www.opticsinfobase.org/submit/style/supplementary-materials-optica.cfm}{Supplementary Materials in Optica}. To reference the supplementary document, the statement ``See Supplement 1 for supporting content.'' should appear at the bottom of the manuscript (above the references).

%\bigskip \noindent See \href{link}{Supplement 1} for supporting content.

%\section{References}

% Bibliography
\bibliography{StokesNaturalSceneGen}

%Manual citation list
%\begin{thebibliography}{1}
%\bibitem{Zhang:14}
%Y.~Zhang, S.~Qiao, L.~Sun, Q.~W. Shi, W.~Huang, %L.~Li, and Z.~Yang,
 % \enquote{Photoinduced active terahertz metamaterials with nanostructured
  %vanadium dioxide film deposited by sol-gel method,} Opt. Express \textbf{22},
  %11070--11078 (2014).
 % \input{StokesNaturalSceneGen.bbl}
%\end{thebibliography}

\end{document}